\documentclass[sigconf,screen]{acmart}
\usepackage{enumitem}

\renewcommand\footnotetextcopyrightpermission[1]{}

\AtBeginDocument{%
  \providecommand\BibTeX{{%
    \normalfont B\kern-0.5em{\scshape i\kern-0.25em b}\kern-0.8em\TeX}}}

\setcopyright{acmlicensed}
\copyrightyear{2018}
\acmYear{2018}
\acmDOI{XXXXXXX.XXXXXXX}

\acmConference[Conference acronym 'XX]{Make sure to enter the correct
  conference title from your rights confirmation emai}{June 03--05,
  2018}{Woodstock, NY}
\acmISBN{978-1-4503-XXXX-X/18/06}

\usepackage{tikz}
\usetikzlibrary{shapes.geometric, arrows, positioning}
\begin{document}

\title{Survey for Landing Generative AI in Social and E-commerce Recsys -- the Industry Perspectives}


\author{Da Xu}
\affiliation{%
  \institution{LinkedIn}
  \city{Sunnyvale}
  \state{California}
  \country{USA}}
\email{dxu2@linkedin.com}

\author{Danqing Zhang}
\affiliation{%
  \institution{Amazon}
  \city{Palo Alto}
  \state{California}
  \country{USA}}
\email{danqinz@amazon.com}

\author{Guangyu Yang}
\affiliation{%
  \institution{Tiktok}
  \city{Santa Clara}
  \state{California}
  \country{USA}}
\email{guangyu.yang@tiktok.com}

\author{Bo Yang}
\affiliation{%
  \institution{Amazon}
  \city{Palo Alto}
  \state{California}
  \country{USA}}
\email{byyng@amazon.com}

\author{Shuyuan Xu}
\affiliation{%
  \institution{Rutgers University}
  \city{New Brunswick}
  \state{New Jersey}
  \country{USA}}
\email{shuyuan.xu@rutgers.edu}

\author{Lingling Zheng}
\affiliation{%
 \institution{Microsoft}
 \city{Redmond}
 \state{Washington}
 \country{USA}}
\email{linzheng@microsoft.com}

\author{Cindy Liang}
\affiliation{%
  \institution{LinkedIn}
  \city{Sunnyvale}
  \state{California}
  \country{USA}}
\email{cliang@linkedin.com}


\begin{abstract}
Recently, generative AI (GAI), with their emerging capabilities, have presented unique opportunities for augmenting and revolutionizing industrial recommender systems (Recsys).
Despite growing research efforts at the intersection of these fields, the integration of GAI into industrial Recsys remains in its infancy, largely due to the intricate nature of modern industrial Recsys infrastructure, operations, and product sophistication. 
Drawing upon our experiences in successfully integrating GAI into several major social and e-commerce platforms, this survey aims to comprehensively examine the underlying system and AI foundations, solution frameworks, connections to key research advancements, as well as summarize the practical insights and challenges encountered in the endeavor to integrate GAI into industrial Recsys. As pioneering work in this domain\footnote{Subsequent versions of this manuscript will be released with additional real-world examples and solution details after clearing the review process.}, we hope outline the representative developments of relevant fields, shed lights on practical GAI adoptions in the industry, and motivate future research.
\end{abstract}


\maketitle

\section{Introduction}
\label{sec:introduction}


\textbf{Motivation} -- the last two years have been a thrilling journey for generative AI (GAI) and its emerging capabilities, revolutionizing and reshaping the technology landscape across major domains \cite{bommasani2021opportunities,yang2023dawn}. Recommendation systems (Recsys) have not been left behind, as the field experiences a surge of innovative ideas and research works that leverage GAI to augment or replace existing system components and layers \cite{liu2023pre,wu2023survey,fan2023recommender}. 
Nonetheless, landing these innovations in the real world systems, especially those for social and e-commerce recommendations \cite{tang2013social,wei2007survey}, poses significant challenges. 
In particular, industrial Recsys are powered by sophisticated and compound AI systems that encompass not only AI models but also infrastructure, operational processes, as well as business and product considerations. Effectively integrating new technologies requires a holistic approach in practice.

Our work emerges at a crucial juncture to provide an \emph{up-to-date}, \emph{application-centric}, and \emph{interdisciplinary} survey tailored for the industry and Recsys community.

\subsection{Outline}

\begin{figure}[htb]
    \centering
    \includegraphics[width=\linewidth]{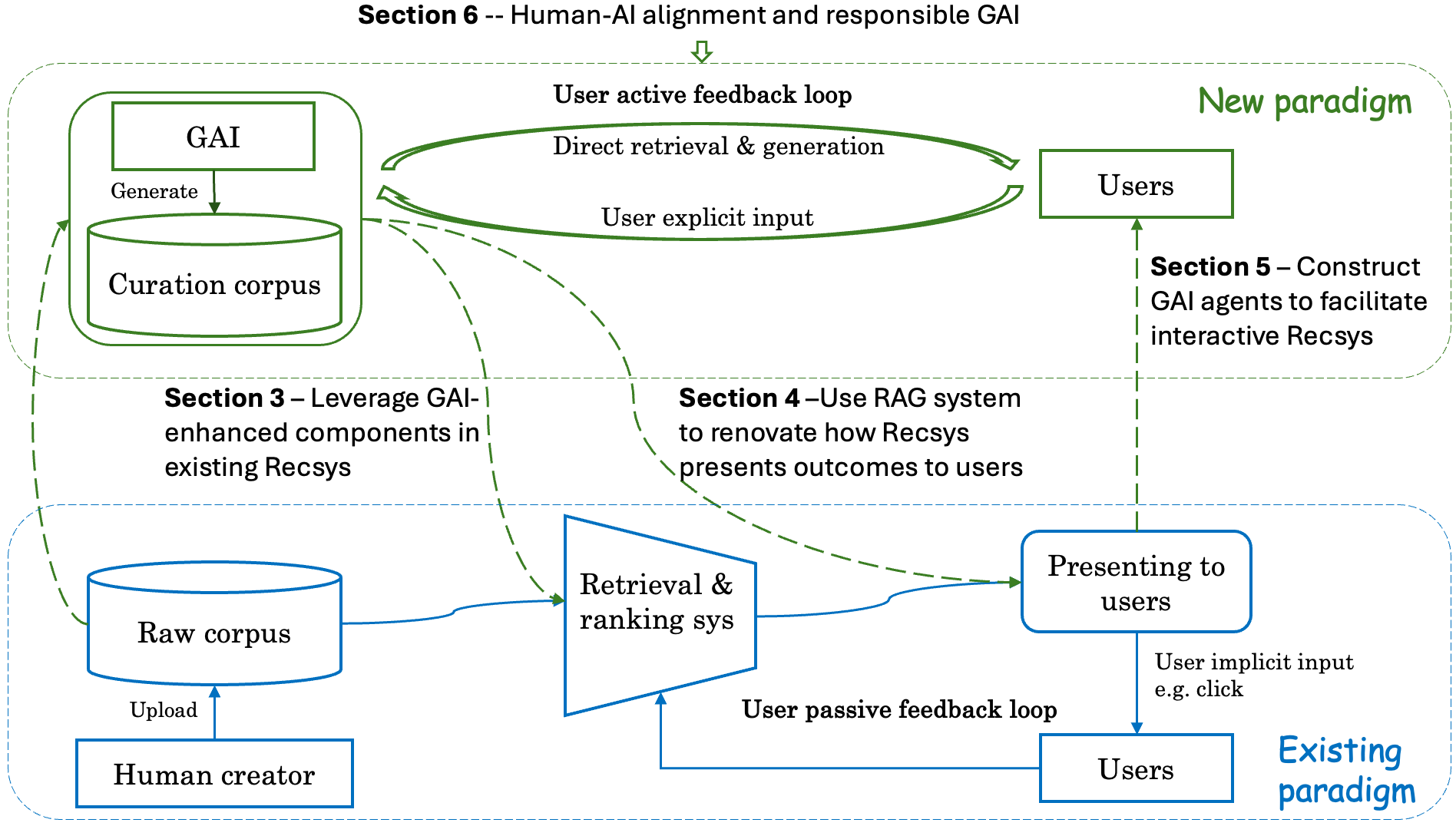}
    \caption{Structure of the survey.}
    \label{fig:outline}
\end{figure}

The first part of the survey aims to offer a concise overview of existing industrial recsys and several key deficiencies, and the major advancements in GAI with a specific focus on GAI production fundamentals and LLMOps. Expanding upon these fundamentals, the structure of our survey can be depicted as in Figure \ref{fig:outline}. 

In Section \ref{sec:personalization}, we start with mapping the opportunity landscape of using GAI usecases to enhance \textbf{personalized recommendation} -- the cornerstone of user satisfaction in the existing Recsys paradigm \cite{ricci2010introduction}. 
After exploring the opportunities and initiatives, we address practical considerations and solutions for integrating the promising ones into real-world production systems.

In Section \ref{sec:curation}, we delve deeper beyond personalized retrieval and ranking, focusing on the utilization of GAI for \textbf{Recsys curation}. This new landscape entails such as re-purposing raw contents, curating from external knowledge, and generating explanations to precisely address users' varied information need and enhance transparency and trustworthiness -- the other crucial dimensions of user satisfaction in social and e-commerce Recsys \cite{nilashi2016recommendation,hassan2019trust,wang2022trustworthy}. Drawing from our experience in deploying these strategies on real-world platforms, we present systematic approaches to facilitate \textbf{retrieval-augmented generation} (RAG) within Recsys accompanied by detailed solutions pertaining to AI modeling, serving, LLMOps, and other practical considerations. 

Section \ref{sec:agent} is devoted to equipping Recsys with \textbf{AI agents} to facilitate \textbf{interactive recommendation} and \textbf{active feedback} loops, marking a departure from the prevailing passive feedback paradigm. By harnessing more advanced tool-using and sequential reasoning-acting capabilities \cite{wang2023survey,yao2022react}, Recsys agents can be engineered to not only suggest items and offer content curation and explanation, but also processing and responding to the explicit, free-form user asks and preference statements. This approach serves to bridge existing gaps and potentially exceed the capabilities and user experiences offered by contemporary conversational Recsys \cite{jannach2021survey}. 


In Section \ref{sec:alignment}, we review two pivotal challenges associated with deploying GAI in customer-facing applications: \textbf{responsible GAI} and \textbf{human-AI alignment} \cite{kenton2021alignment,zou2023universal}. We present the strategies employed by industry applications to navigate the intricate landscape and address the multifaceted aspects of these challenges.
Subsequently, in Section \ref{sec:open-problem}, we summarize and discuss the practical challenges and problems that have surfaced during our endeavors to land GAI in Recsys.

\subsection{Key Contributions and Distinguishing from Related Works}

While several academic surveys have been conducted on the research intersection between LLM/GAI and Recsys \cite{liu2023pre,lin2023can,wu2023survey,fan2023recommender}, there has \emph{not} yet been a review addressing the integration and development of GAI within real-world Recsys. This survey work goes beyond previous works by not only covering new aspects such as \emph{system architecture}, \emph{LLMOps}, \emph{GAI production framework design}, but also establishes links between practical application and a broader spectrum of GAI topics such as human-AI alignment and responsible AI practices. Additionally, we explore the unresolved challenges faced in real-world practice outside the confines of research labs.

\section{Background and Fundamentals}
\label{sec:overview}
This section aims to offer a concise overview of the foundational components of industrial Recsys and the merging GAI ecosystem, with an emphasis on the practical \emph{operational} aspects inherent in real-world compound AI systems.

\subsection{Industrial Recsys in a Nutshell}

\begin{figure}[hbt]
    \centering
    \includegraphics[width=1\linewidth]{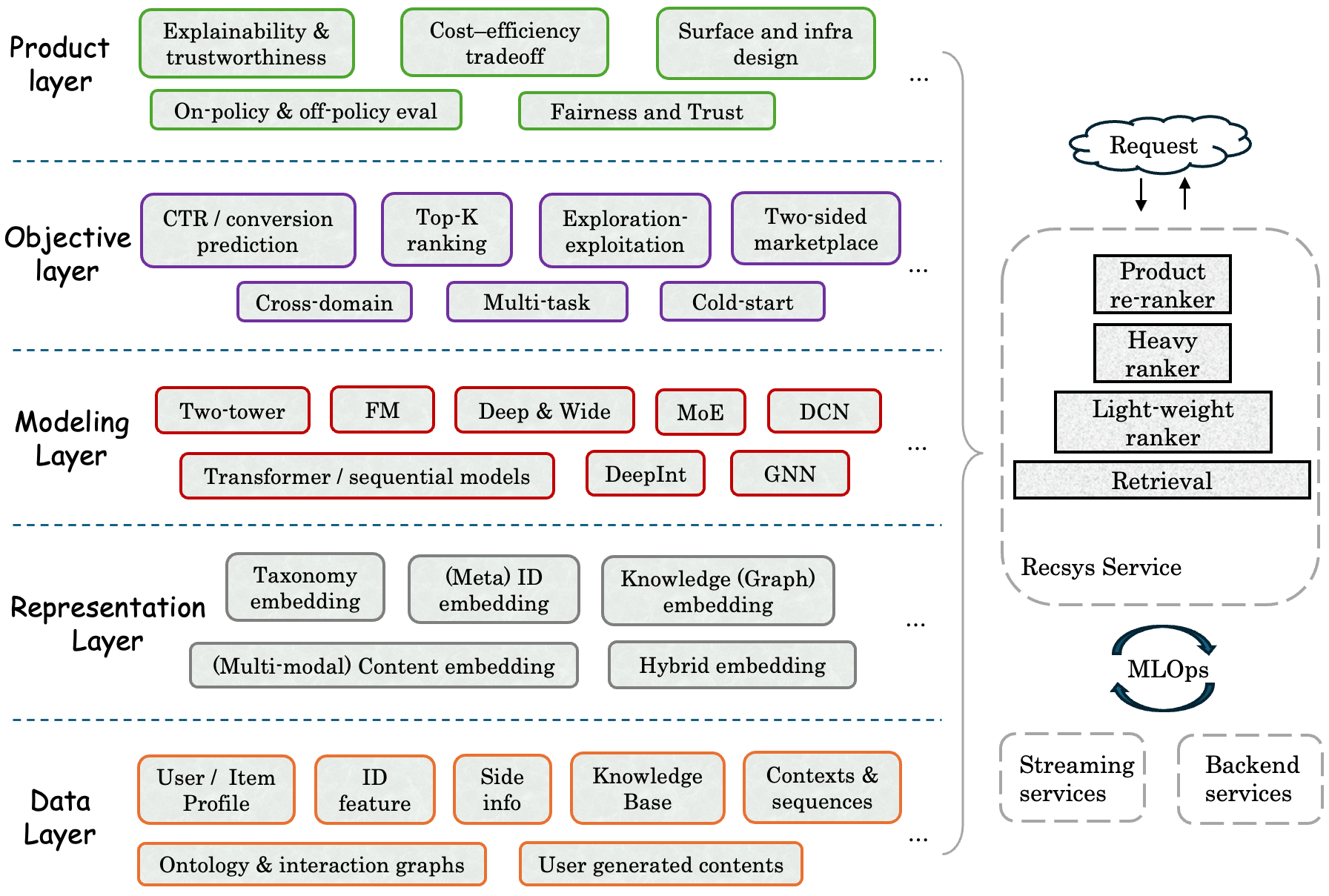}
    \caption{Overview of the key components employed by modern industrial social and e-commerce Recsys. The modeling stack typically consists of data, representation, modeling, objective, and product layers. The infra stack often comprises diverse serving, streaming, backend serveries orchestrated using MLOps techniques.}
    \label{fig:recsys}
\end{figure}

For all major social and e-commerce platforms, user satisfaction and commercial value growth often hinge upon the efficacy of Recsys in meeting increasingly challenging business goals \cite{aggarwal2016recommender,naumov2019deep,ying2018graph,zhang2019deep}. As we illustrate in Figure \ref{fig:recsys}, modern industrial Recsys have evolved into \emph{compound AI systems}, comprising of multiple components that exhibit intricate interplay. The necessity of using compound AI systems stems from several factors.
\begin{itemize}[leftmargin=*]
    \item The business and performance goals of industrial Recsys are \emph{dynamic} and \emph{vary widely}. Compound AI systems can adapt quickly due to their modular structure and operational flexibility.
    \item For a wide array of real-world tasks, constructing components and managing their interplay within the existing system often lead to \emph{simpler} and more \emph{cost-effective} solutions.
    \item Compound AI systems offer \emph{finer levels of control} throughout the product lifecycle, including development, integration, testing, trouble shooting, and maintenance.
\end{itemize}

As subsequently demonstrated in this survey, incorporating GAI solutions also necessitates a holistic approach with \emph{system thinking}. This entails considering the current system architecture, infrastructural limitations, operational capabilities, and product/service-level agreements to avoid unintended consequences. Conversely, GAI solutions can capitalize on the mature solutions including reusable components and design patterns inherent in existing Recsys.

\subsection{Major Deficiencies of Existing Systems}

GAI and foundation models can effectively address several major Deficiencies presented in many existing industrial Recsys:

\begin{itemize}[leftmargin=*]
    \item \textbf{Insufficient open-world memorization and generalization} -- the memorization and generalization (which are the foundations of personalization) of most social and e-comm Recsys relies primarily on members' in-app/web data and engagements.
    \item \textbf{Lack of active content curation} -- industrial social and e-comm Recsys are typically designed to passively retrieve and rank raw contents, with minimum focus on repurposing and curating contents to meet user's highly diversified information indeed. 
    \item \textbf{Absence of interactive reasoning} -- the majority of industrial Recsys operate within the one-shot engagement paradigm and are unable to facilitate interactive reasoning regarding the scene, context, user need, explanation, etc.
\end{itemize}

\subsection{Building GAI Foundation and LLMOps}

\begin{figure}[hbt]
    \centering
    \includegraphics[width=\linewidth]{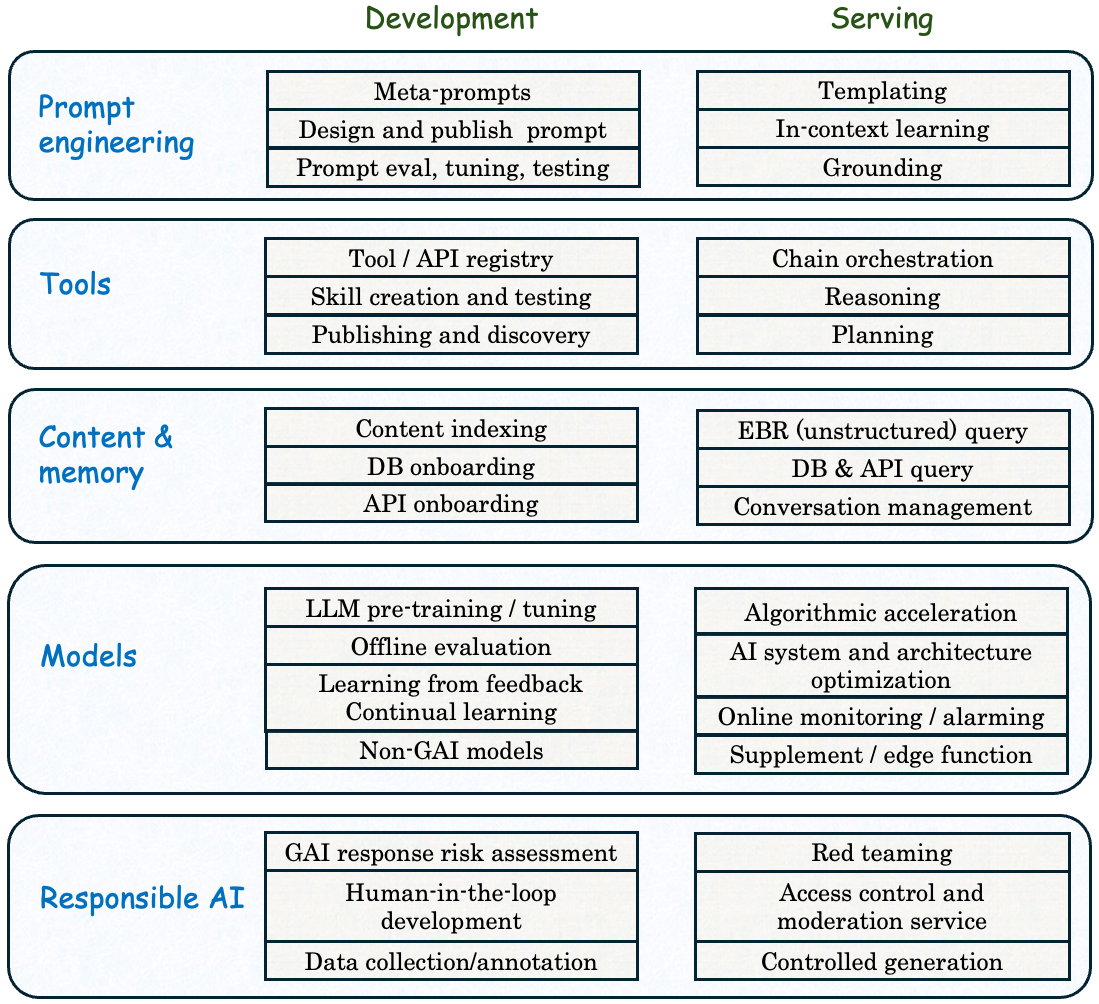}
    \caption{Illustration of the GAI foundations that will be the focus areas of our tutorial.}
    \label{fig:gaiecosystem}
\end{figure}
\begin{figure}[hbt]
    \centering
    \includegraphics[width=0.8\linewidth]{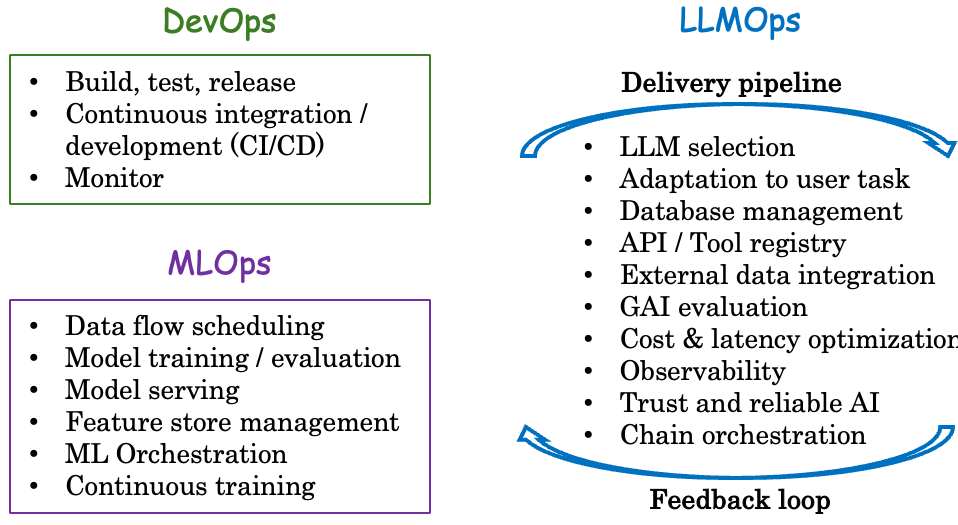}
    \caption{Overview of DevOps, MLOps, and LLMOps. The key LLMOps terms will be clarified in the rest of this survey.}
    \label{fig:llmops}
\end{figure}

Developing and serving GAI can involve \emph{modeling}, \emph{infrastructure}, and \emph{operational} components that are not readily available in Recsys. We provide a comprehensive overview in Figure \ref{fig:gaiecosystem} and \ref{fig:llmops} that is in accordance with the scope of this tutorial.

\subsubsection{GAI ecosystem.} 
Prompt engineering is an effective approach to direct LLM towards producing desired outcome for specific tasks.
Prompting LLM and applying relevant techniques like in-context learning, chain-of-thought, and other enhancements has rapidly become essential for adopting LLM for specific tasks \cite{wei2022chain,liu2023pre,min2022rethinking,zhao2021calibrate,wang2022self}. Figure \ref{fig:gaiecosystem} provides a summary of the elements required for implementing prompt engineering solutions in production, which we will elucidate throughout the remainder of this survey.

Prompting LLMs can be further endowed with \emph{tool-using} capabilities \cite{schick2024toolformer,yang2024gpt4tools}, enabling them to address compositional tasks through chains of \emph{reasoning and action} \cite{yao2022react}. This also catalyzed the development of LLM-based autonomous agents that integrate planning, memory (including content comprehension), and tool utilization \cite{wang2023survey,xi2023rise, bubeck2023sparks}. In Figure \ref{fig:gaiecosystem}, we illustrate the requisite services and functions for accessing the internal and external services, tools, and memory within production systems. 

While prompt engineering has demonstrated reasonable proof-of-concept results across various tasks, scaling up and ensuring sustainability of GAI solutions often requires \textbf{GAI model optimization}, which is a key direction to address cost, latency, and the effective usage of proprietary data and domain knowledge. Fortunately, task-specific GAI model development is often simpler through effective fine-tuning \cite{hu2021lora,li2021prefix} and learning from human feedback \cite{ouyang2022training,rafailov2024direct}. Training GAI solutions from scratch is often less undesired \cite{brown2020language,raffel2020exploring}. Serving up GAI deployment usually involves a collaborative refinement of the algorithm (e.g. through model compression, sparsity, operator fusion, distillation, pruning, efficient decoding techniques \cite{zafrir2019q8bert,jiao2019tinybert,pope2023efficiently,niu2021dnnfusion}) and the AI system architecture (e.g. with quantization, PagedAttention, parallel computation \cite{kwon2023efficient,miao2023towards}). Finally, we mention that non-GAI models will remain essential in the ecosystem as \emph{supplement and edge functions}.



\subsubsection{LLMOps.} Similar to other internet applications, delivering GAI products that can meet client expectations remains a significant challenge unless the workflows can be effectively automated and operationalized. This is addressed within the domain of LLMOps (as we outlined in Figure \ref{fig:llmops}), which represents an evolution of \emph{development operations} (DevOps) and \emph{machine learning operations} (MLOps) \cite{kreuzberger2023machine,jabbari2016devops}. 
In addition to leveraging insights from the previous generation of software and machine learning product development, there are several new practices and concepts crucial for harnessing and managing the power of GAI. The most critical ones include \emph{GAI evaluation} and the \emph{trust and responsible GAI} \cite{Baxter2023risk,chang2023survey}, which will be further elaborated in the next sections.


\subsection{Classification Framework of GAI in Recsys}

\begin{figure}[hbt]
    \centering
    \includegraphics[width=\linewidth]{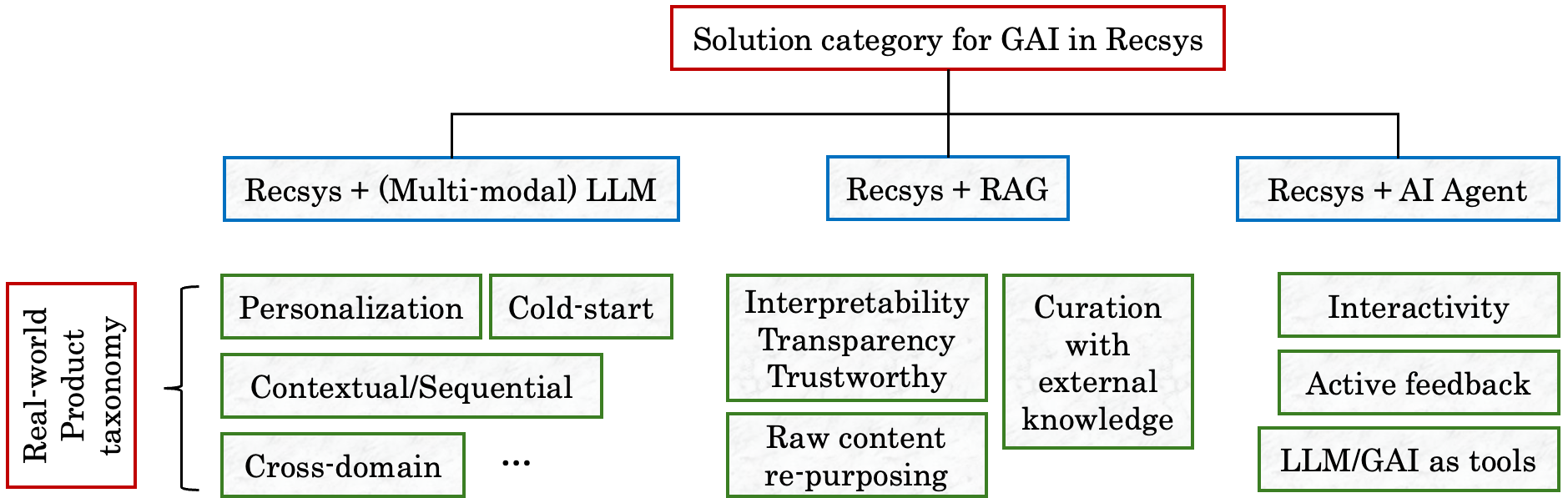}
    \caption{Real-world product taxonomy of social/e-commerce Recsys and they best fit into the major solution categories of GAI in Recsys solutions.}
    \label{fig:taxonomy}
\end{figure}

Emphasizing \emph{practicality} and \emph{feasibility} in real-world deployment, our taxonomy and classification framework (depicted in Figure \ref{fig:taxonomy}) is organized by: 1). real-world production tasks of social and e-commerce Recsys, 2). their \emph{optimal realization} through three primary categories of GAI methods.
Notably, our framework is \emph{application-centric}, thus facilitating a comprehensive examination of the deployment of GAI in real-world Recsys.

\section{GAI-enhanced data, feature, and modeling for Recsys}
\label{sec:personalization}





For most existing social and e-commerce Recsys, the goal of recommendation is to identify good matches and tailor contents and experiences to best align with \emph{personalized preferences} \cite{ricci2010introduction}. 
Given GAI's unprecedented capability to process complex multi-modal user-item information/contexts/sequences \cite{yuan2023go,li2023text,sarkar2023outfittransformer}, solve cold-start/few-shot problems \cite{ding2021zero,hou2024large}, and serve as various other components outlined in Figure \ref{fig:recsys}, several preliminary studies have recognized the potential of GAI to augment personalized recommendation.

\subsection{Research Progress}

\begin{figure}[htb]
    \centering
    \includegraphics[width=\linewidth]{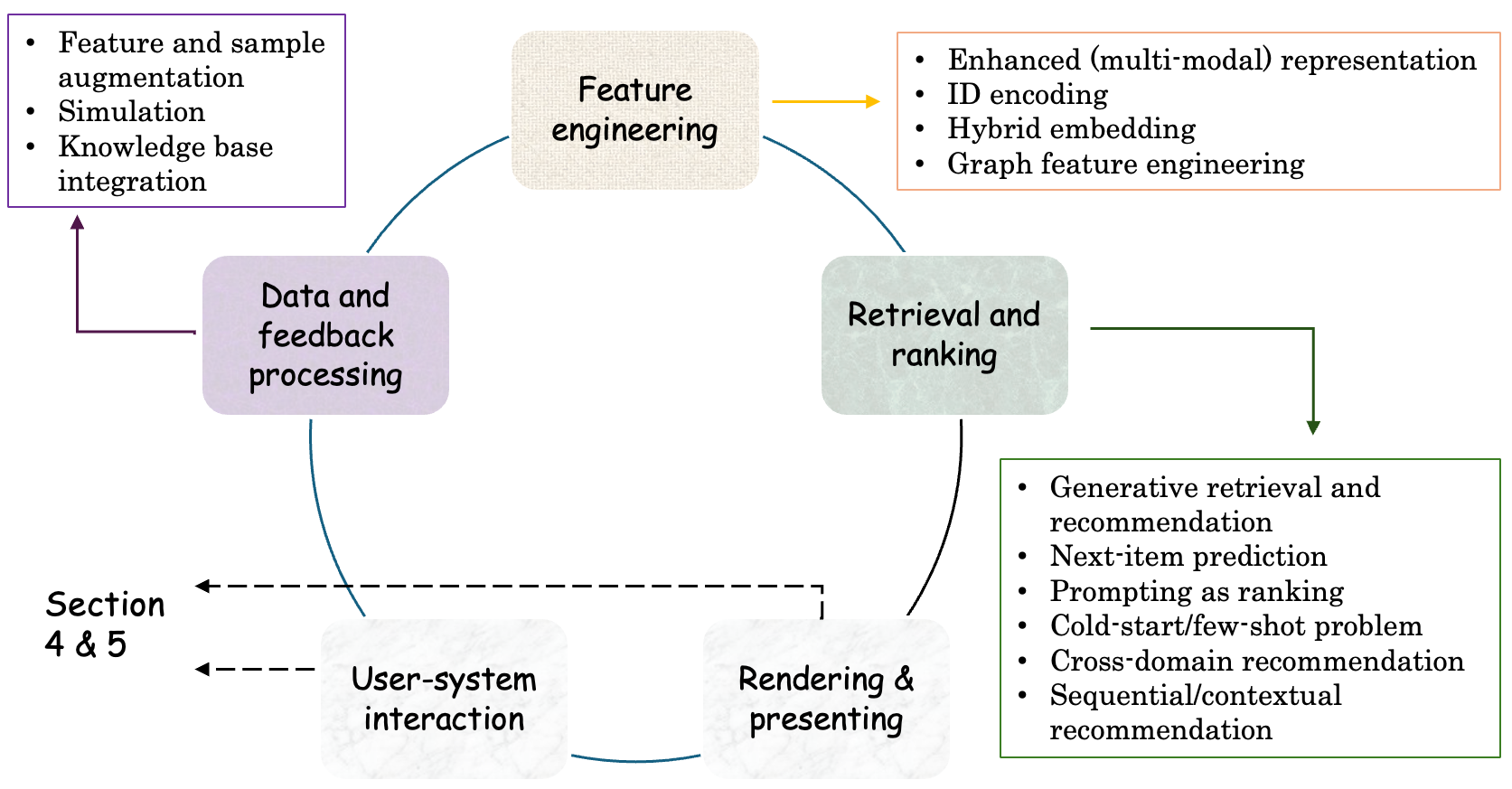}
    \caption{Overview of the predominant trends of using GAI to improve personalized retrieval and ranking in social and e-commerce Recsys. We also include  rendering and interaction stages in the circle for completeness.}
    \label{fig:GAIinRecsys}
\end{figure}

In Figure \ref{fig:GAIinRecsys}, we present a comprehensive overview of the predominant trends in existing efforts, where the \emph{data and feedback processing} and \emph{feature engineering} stages are also taken into account. 
The key advancements in applying GAI for data and feedback processing include enriching user/item profile extraction and tagging \cite{li2023taggpt,brinkmann2023product}, comprehending user intent and interests \cite{christakopoulou2023large}, condensing and augmenting data \cite{wu2023leveraging}, integrating external knowledge \cite{chen2023knowledge,yin2023heterogeneous}, and simulating and generating records \cite{wang2023recagent}. We note that there is a growing body of literature on using Agent as Recsys user simulator \cite{zhang2023generative,zhang2023agentcf,wang2023recagent}, which represents an alternative application of agents to the one we will present in Section \ref{sec:agent}.

For \emph{feature engineering}, Recsys primarily utilize sparse categorical features and their representations. Through not entirely GAI-driven, the advent of LLM has presented opportunities to further improve representation learning especially for textual features and multi-modal data \cite{zhu2023collaborative,ren2023representation,sarkar2023outfittransformer}. Graph data has also been introduced to enhance LLM's capability to generate high-quality representations \cite{wei2024llmrec,xie2023graph}. Recent studies have proposed using LLM for ID-based representation of users and items \cite{geng2022recommendation,hua2023index}. 

GAI techniques have also been explored for \emph{retrieval and ranking tasks}. Generative retrieval and recommendation has emerged as a promising domain where GAI directly generates item IDs as output \cite{sun2024learning,rajput2024recommender,xu2023openp5}. Some studies also follow NLP approaches to use GAI for top-K recommendation \cite{dai2023uncovering}. Other major directions include context-aware recommendation where GAI's world knowledge can serve as important background information \cite{xi2023towards,harte2023leveraging}. For a more detailed examination of this subject, please refer to the comprehensive surveys of \citet{lin2023can} and \citet{wu2023survey}.

\subsection{Solution Framework and Practical Considerations}


Since the subject matter addressed in this section usually represent the initial attempts of adopting GAI to the offline development of real-world Recsys, investing in GAI foundation and LLMOps (especially the offline elements described in Figure \ref{fig:gaiecosystem} and \ref{fig:llmops}) can substantially expedite GAI-in-the-loop endeavors in the future. During this stage, latency is typically not a significant obstacle for batch offline inference, and prompt engineering/fine-tuning LLM can often deliver tangible improvements across various tasks \cite{liu2023pre}. Furthermore, the existing metrics and frameworks for offline and online Recsys evaluation are still applicable \cite{castells2022offline,kohavi2020trustworthy}, as the final forms of recommendation remains unchanged.

\begin{figure}[htb]
    \centering
    \includegraphics[width=\linewidth]{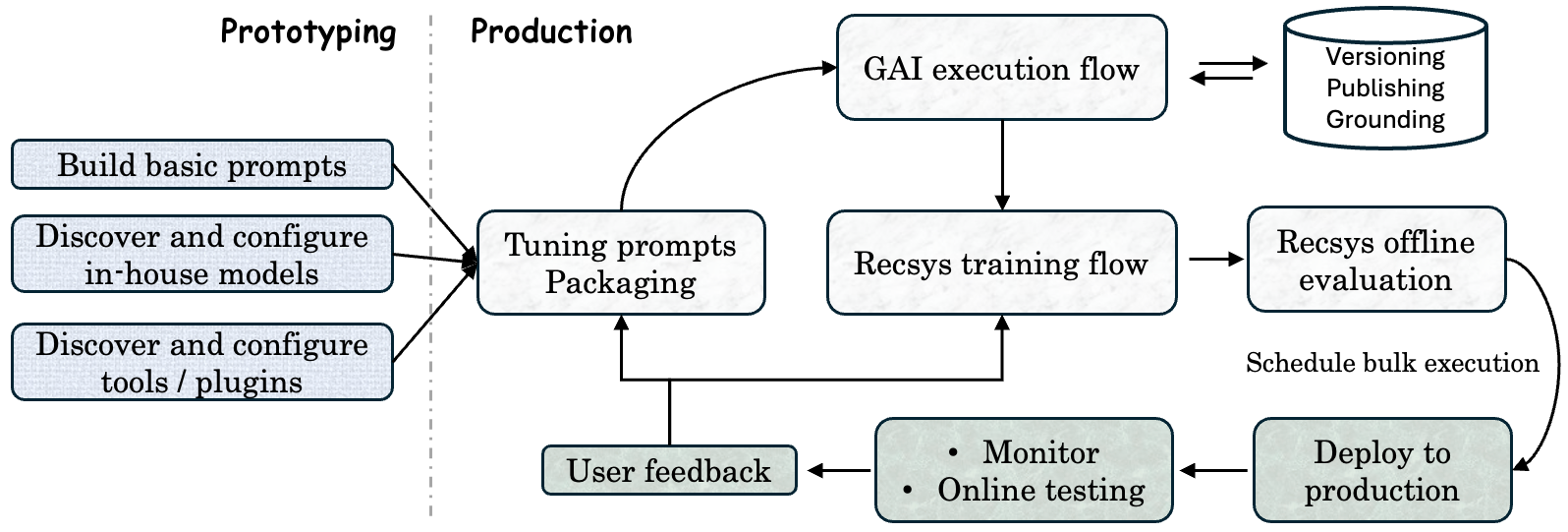}
    \caption{A standard workflow for productionizing prompt engineering solution for enhancing Recsys training.}
    \label{fig:GAI4Recsys-flow}
\end{figure}

Nevertheless, while leveraging prompt engineering for the development of novel features may seem straightforward, ensuring the seamless transition of temporary prototypes to large-scale production necessitates \emph{operational strategies} (see Figure \ref{fig:GAI4Recsys-flow}). This process entails the construction of an end-to-end pipeline that facilitates effectively prompt tuning and evaluation \cite{liu2021p}, versioning and publishing prompts \cite{wu2023framework}, and grounding and monitoring to ensure the correct services and contexts are triggered and utilized. 

In practice, we have observed that using "small" in-house and task-specific GAI models can offer cost and performance advantage over using the open "large-model" solutions. 
This finding is substantiated by the fact that many smaller-scale applications usually do not require GAI excelling at both task specialization and generation \cite{schick2020s,bommasani2021opportunities}. Nevertheless, ensuring the efficacy of "small" models often require model fine-tuning and alignment with proprietary data, as well as the development of instruction, demonstration, and in-context learning strategies \cite{liu2023pre,min2022rethinking,liu2021p}. These important aspects are also revealed from the recent industrial efforts \cite{cui2022m6,fan2023recommender}.

Lastly, we note that personalized recommendation has been extensively studied in the past two decades, leading to many practical and effective domain methods \cite{perugini2002recommendation,ko2022survey}. 
Considering the maturity and efficacy of many time-tested industrial Recsys,
the potential value of integrating GAI may be more pronounced in domains where existing RecSys exhibit limitations. These areas will be explored in detail in the subsequent sections.

\section{Augmenting the Curation Capability of Recsys}
\label{sec:curation}
Beyond enhancing personalized retrieval and ranking, GAI-powered curator allows Recsys to transcend the limitation of only showing human-generated contents. It enables \textbf{re-purposing}, \textbf{explaining}, and \textbf{curating contents using external knowledge} to meet diversified needs and elevate transparency and trust \cite{wang2023enhancing,wang2023generative,wang2022trustworthy}. 

However, using GAI to directly generate customer-facing contents can suffer from \emph{inadequate knowledge} of both the user and subject matter, \emph{limited task expertise}, \emph{lack of control of the output}, and various other \emph{trust and safety issues} that cannot be fully addressed through improving prompting or model tuning techniques \cite{huang2023survey,petroni2019language,carlini2021extracting}.
As our earlier discussions suggest, building compound AI system can be a pragmatic approach to systematically address the above challenges, which leads to our introduction of the \emph{retrieval-augmented generation} (RAG) -- a technique that integrates external data retrieval into the generative process \cite{li2022survey,lewis2020retrieval}.

\subsection{Research Progress}

\begin{figure}[hbt]
    \centering
    \includegraphics[width=\linewidth]{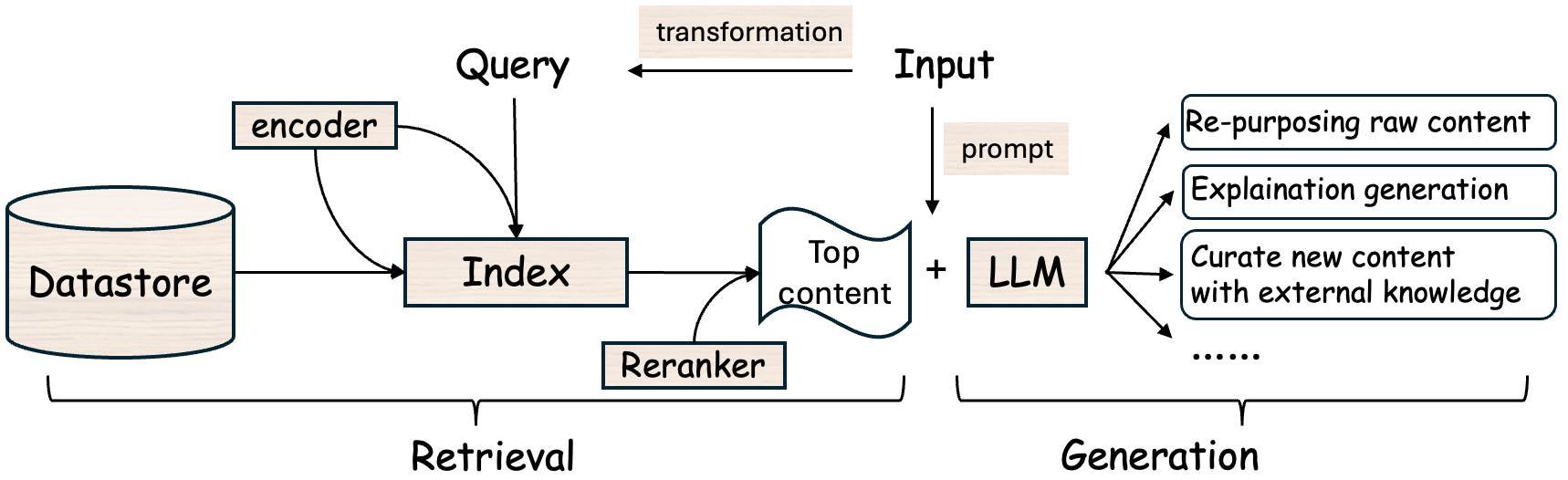}
    \caption{Illustration of a typical RAG system marked with the areas that can be optimization within the system.}
    \label{fig:rag-naive}
\end{figure}

Until not long ago, Recsys' curation capability (e.g. explanation generation) primarily relies on the efficacy of templates and manual curation \cite{li2020generate}, which often lack richness and in-depth context comprehension and knowledge. 
RAG systems combine the strengths of retrieval-based and GAI-based methods to enhance the accuracy, credibility, and relevance of the GAI outputs for various Recsys usages \cite{di2023retrieval}. The scope of retrieval can extend beyond user-generated content to, for example, behavioral sequences \cite{lin2023rella}. We provide a concise overview of RAG system in Figure \ref{fig:rag-naive}.

RAG is a rapidly developing technique (with variants such as self-RAG, auto-RAG, Corrective RAG) where the workflow is further improved by adding components like retrieval result verification and re-ranking \cite{yan2024corrective, zhuang2023open, li2023llatrieval}, adaptive retrieval \cite{jiang2023active}, and query rewrite \cite{ma2023query}. 
Typically, incoming requests are first parsed into queries and prompts. The queries are directed to the retrieval and re-ranking service (with pre-trained encoder and datastore index) to retrieve the most relevant contents. The contents are subsequently leveraged to improve generation by such as \emph{enriching/augmenting} the input prompts and \emph{correcting} LLM's output.


It is noteworthy that the output of RAG can be \emph{multi-modal}, including images (including attribute change of existing images and stylized images) and videos, through the integration of diffusion and CLIP models \cite{rombach2022high,radford2021learning}. The \emph{adaptability} and \emph{versatility} of RAG have positioned it as a prominent GAI solution, particularly suited for tasks requiring integration with excessive knowledge. \cite{gao2023retrieval}. As we will demonstrate in the subsequent section, RAG also holds promise for effective utilization by AI agents.

\subsection{Solutions Framework and Practical Considerations}
\label{sec:RAG-solution}

\begin{figure*}[tbh]
    \centering
    \includegraphics[width=0.9\linewidth]{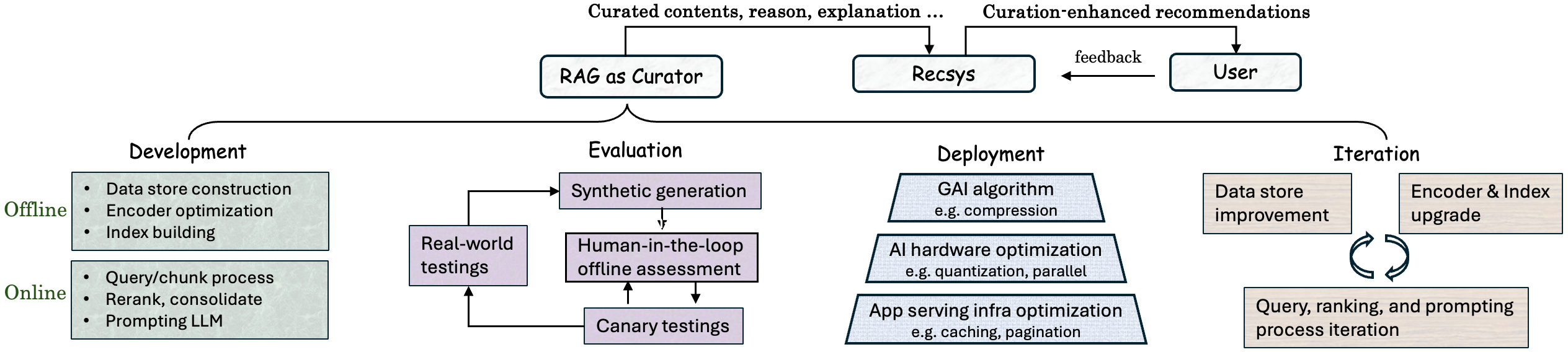}
    \caption{Taxonomies and overview of the solution framework for building RAG as GAI curator in Recsys.}
    \label{fig:RAG-landing}
\end{figure*}

Despite that RAG shares several common building blocks with Recsys and holds promising benchmark results \cite{chen2023benchmarking}, implementing and integrating RAG into Recsys presents significant challenges in practice. We summarize them into three folds as below, and present an overview of the solution framework in Figure \ref{fig:RAG-landing}. Generally, addressing these challenges requires system-level solutions.

\begin{enumerate}[leftmargin=*]
    \item \textbf{evaluation} -- unlike evaluating item recommendations, assessing curation outcome in Recsys is often feasible only during runtime, and the metric taxonomy for \emph{natural language generation} (NLG) differs from that of RecSys \cite{sai2022survey,zhou2022deconstructing};
    \item \textbf{deployment} -- deploying RAG within the service-level agreement (SLA) of Recsys necessitate optimizations across algorithm, AI system, and the serving infrastructure;
    \item \textbf{building and refining RAG} involve fostering synergies among multiple interacting components, both offline and online. 
\end{enumerate}

\subsubsection{Evaluating RAG performance offline.} RAG system is difficult to evaluate offline because of the dependency on real-time and operation contexts. We have found the application of \emph{canary testing}, a technique derived from DevOps practices \cite{tarvo2015canaryadvisor}, along with synthetic generation to be effective. Canary testing enables the real-time performance tracking of RAG by replicating online requests within the same context without impacting user experience. Leveraging the tracked events, we can synthesize records and conduct various NLG evaluations, such as employing BLEU, ROUGE, and other hallucination metrics \cite{ji2023survey,gatt2018survey}, and performing risk assessment (refer to Section \ref{sec:alignment} for further detail).

In contrast to traditional NLG tasks like translation, summarization, and open-book Q\&A, there lacks well-established evaluation solutions for such as \emph{explanation generation} for Recsys. Our practical experience suggests that expert evaluation, even on a modest scale, proves highly effective. Nonetheless, it is crucial to establish explicit standards and guidelines to minimize ambiguities and misunderstandings during expert evaluation \cite{howcroft2020twenty}. Lastly, standard relevance metrics such as precision@k and recall@k can be helpful for specific tasks where ground truth is available.

\subsubsection{Development.} For \textbf{building} the initial RAG solution, a typical roadmap involves two phases: 
\begin{itemize}[leftmargin=*]
    \item \textbf{offline} -- constructing data stores along with the development of the encoder and index for each data store \cite{karpukhin2020dense};
    \item \textbf{online} -- setting up execution flows for query processing, chunk re-rank/process/consolidate, and prompting LLM \cite{guu2020retrieval,borgeaud2022improving}. 
\end{itemize}
Following these phases, the RAG system can be encapsulated as a standalone service module, responsible for managing incoming requests from the RecSys (as illustrated in Figure \ref{fig:RAG-landing}). 

\subsubsection{Deployment.} Given that social and e-commerce Recsys often adhere to stringent SLA \cite{kersbergen2021learnings}, may encounter challenges in meeting these SLAs when handling inputs or outputs of moderate lengths. As elaborated in Section \ref{sec:overview}, optimizing LLM inference typically involves a combination of algorithmic and AI system enhancements \cite{miao2023towards}. Each of these methods entails its own trade-offs, so their efficacy should be discussed case-by-case. In practice, we have found adopting traditional web service optimization techniques, such as dynamic caching and pagination, proves valuable for circumventing the SLA \cite{cao1998active,fredrich2012restful}. Furthermore, these techniques have been adapted to specifically support LLM applications  \cite{bang2023gptcache}.

\subsubsection{Iteration.} Iterating and refining RAG solution post-launch often involve encoder and index update through sequential and asyncornized/batch training, improving the data store, and optimizing the querying process, chunk processing logic, and prompting strategies \cite{izacard2022atlas}. These components are also highlighted in Figure \ref{fig:rag-naive}. 

However, it is noteworthy that directly optimizing RAG based on \emph{implicit} feedback from the recommendations, such as clicks, can be challenging due to the feedback being attributed to a combination of raw and curated contents. 
Consequently, establishing explicit and active feedback loops for \emph{richer} user feedback can benefit the optimization of GAI curator in Recsys. This topic will be further explored in the next two sections.

\section{Facilitating Action and Reasoning for Interactive Recsys}
\label{sec:agent}
Within the current Recsys paradigm, user engagement is generally facilitated via single-round, passive interactions through a constrained set of predefined mechanisms, including item icon clicks or search bar queries \cite{aggarwal2016recommender,ricci2010introduction}. This framework is tailored for scenarios in which passive feedback is considered sufficient for both user satisfaction and business objectives.

However, such Recsys often lack ability to drive active interactions or leverage external tools (e.g. to access external knowledge bases) to further engage and satisfy users. Consequently, their applicability is restricted, and cannot generalize across recommendation scenarios or to meet emerging user demands for \textbf{interactivity} and participating in \textbf{active feedback} loops.  
To address these shortcomings, we advocates for the augmenting RecSys with autonomous AI agent, harnessing their robust natural language reasoning capabilities and tool utilization proficiency \cite{wang2023survey,xi2023rise}, in conjunction with the GAI solution frameworks discussed in Section \ref{sec:personalization} and \ref{sec:curation}.

\subsection{Research Progress}

From Section \ref{sec:personalization} and \ref{sec:curation}, we see that GAI can indeed enable interactive recommendations by curating human-like responses for various tasks such as direct recommendation, preference refinement, recommended follow-up questions, and recommendation justification \cite{jannach2021survey, huang2024foundation}. AI agents can be employed to manage and orchestrate these interactive components \cite{deng2024large, wang2023recmind}. In fact, the RAG system introduced in Section \ref{sec:curation} can be considered a simple AI agent with retrieval and generation modules.

\begin{figure}[htb]
    \centering
    \includegraphics[width=\linewidth]{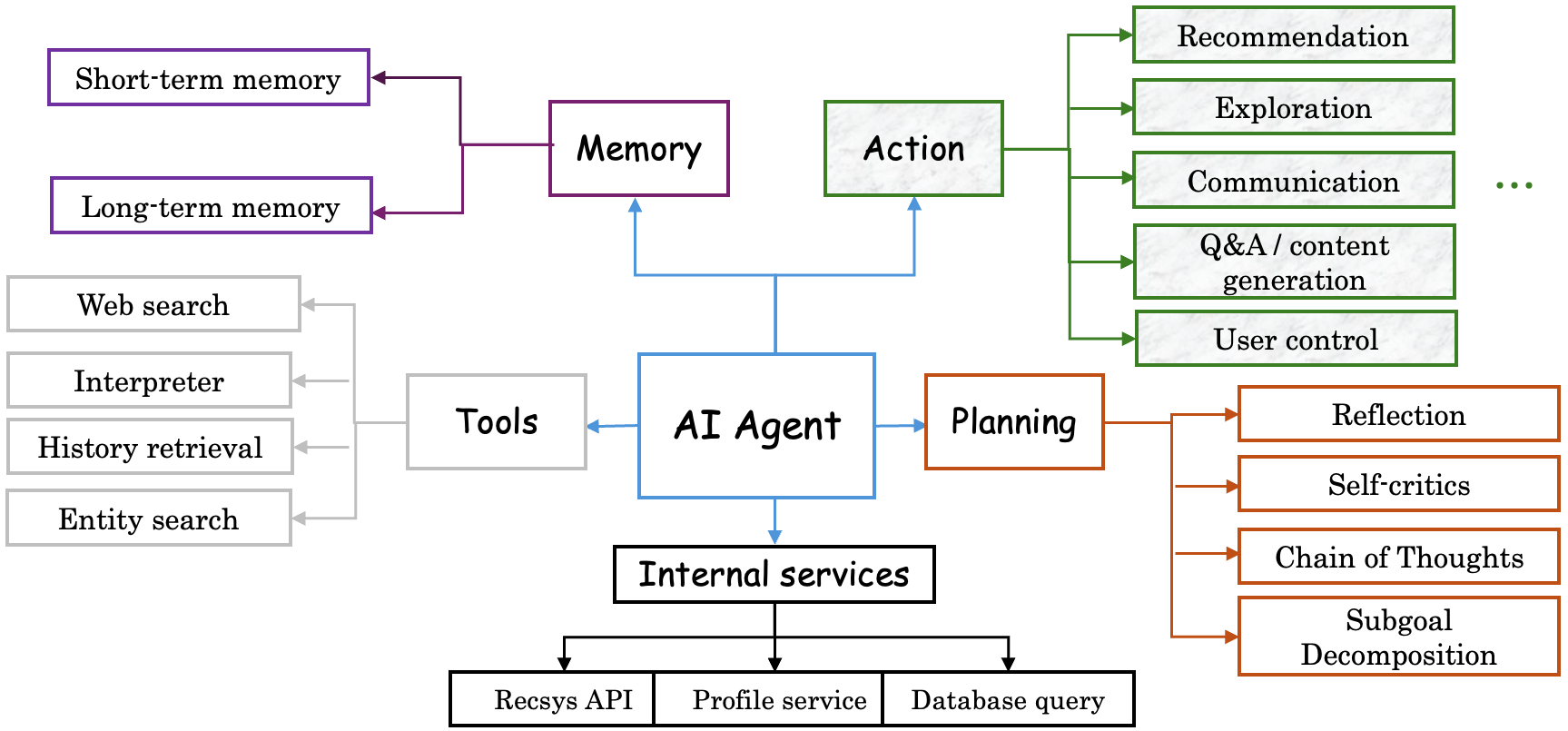}
    \caption{Illustration of the components and capabilities of an AI agent for typical social and e-commerce scenarios.}
    \label{fig:agent-overview}
\end{figure}

LLM-powered AI agents can utilize LLM to reason through problems, create plans, produce text-based outputs and actions, and interact with external tools via API calls.
The \emph{planning} aspect can be facilitated by either providing the agent with a predefined reasoning graph, as discussed in \cite{yang2022seqzero}, by harnessing LLM's reasoning abilities to dynamically generate the reasoning path \cite{wei2022chain}.
For \emph{actions}, the agent can generate textual outputs following the decision-making processes regarding whether and how to engage with a particular tool \cite{schick2024toolformer}. Recent research has delved into both of these capabilities for Recsys \cite{wang2023recmind,huang2023recommender}.

\subsection{Solution Framework and Practical Considerations}

We start by conceptualizing an AI Agent (as shown in Figure \ref{fig:agent-overview}) that is equipped with a diverse array of capabilities to facilitate interactive recommendation processes.
Depending on the product design and requirement, the planning module can be one of the two types, where the main difference lies in how planning is executed: \emph{predefined} or \emph{generated in real time}.

\subsubsection{Agent with pre-defined reasoning graph.}

\begin{figure}[htb]
    \centering
    \includegraphics[width=\linewidth]{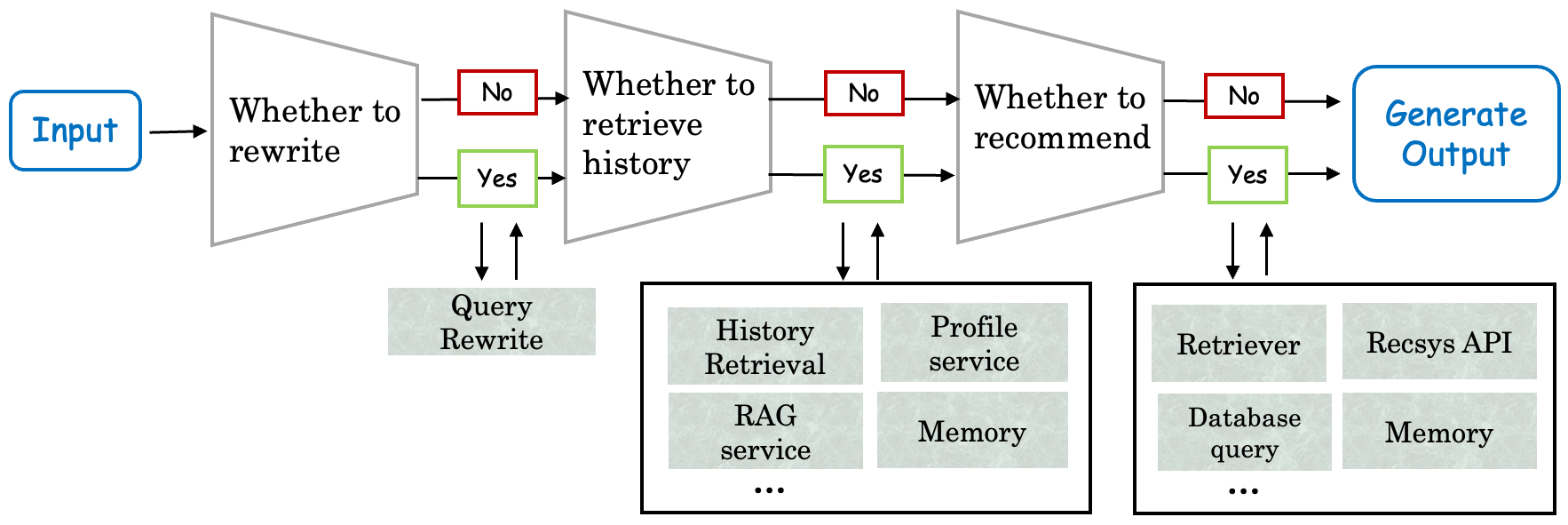}
    \caption{Example of Type-1 GAI agent with pre-defined reasoning graph for in-session recommendation tasks.}
    \label{fig:gai-agent}
\end{figure}

We outline one example in Figure \ref{fig:gai-agent} where the pre-defined reasoning graph is applied, and the role of LLM is to make decisions on how to navigate the graph and use tools to execute each step along the sequence.
This type of Agent is designed to minimize latency, cost, and \emph{operation overhead} -- instead relying on LLM to generate outputs for a function call, we apply rules and light-weight supplement models for decision-making. However, this approach is mostly suitable for simpler applications with small action spaces. Also, creating a predefined reasoning graph also requires domain expertise, and many not be applicable in certain scenarios.

\subsubsection{Agent that uses GAI for planning.}

\begin{figure}[htb]
    \centering
    \includegraphics[width=\linewidth]{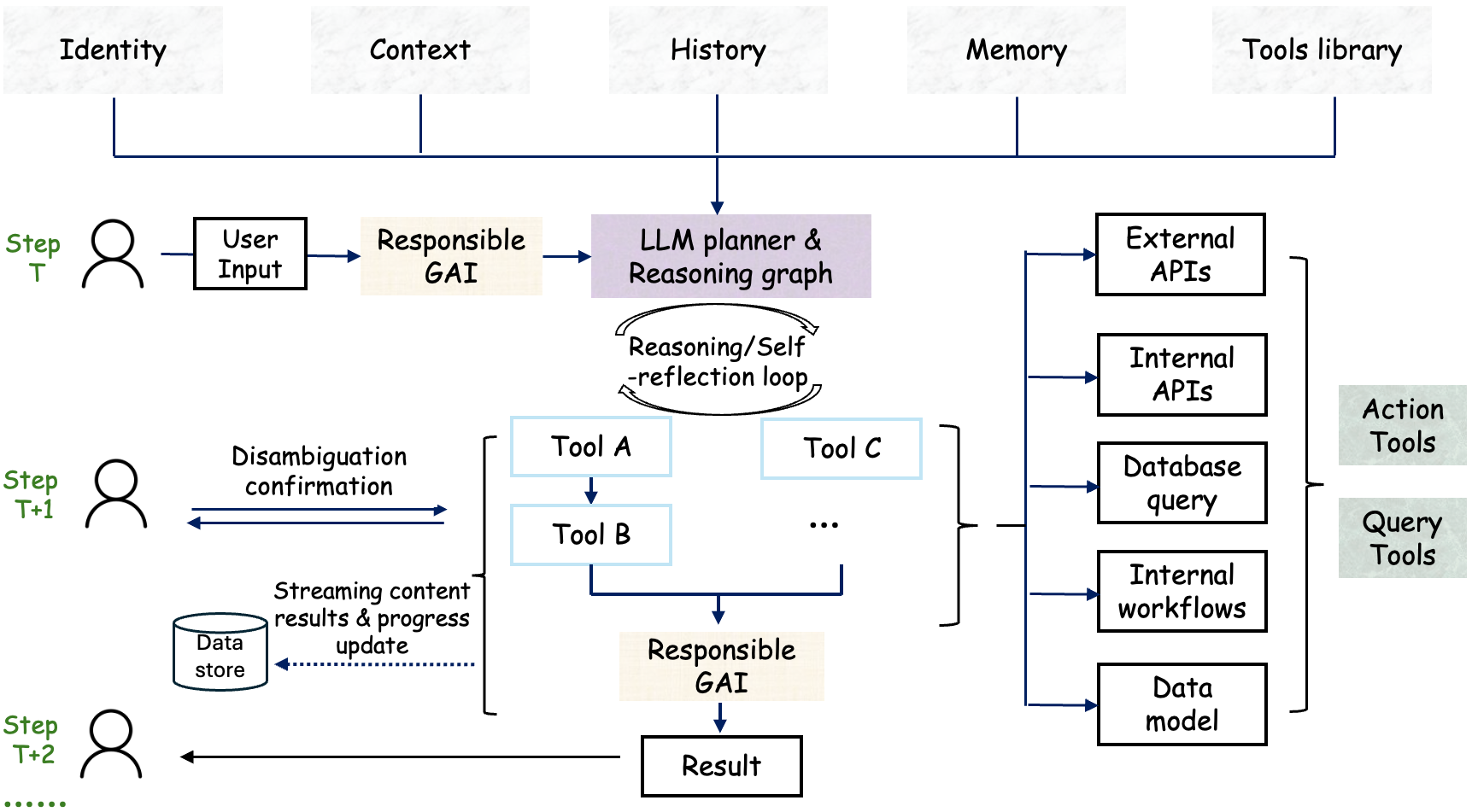}
    \caption{Overview of a real-world Recsys agent for multi-round interactive social recommendation. The Responsible GAI component will be discussed in Section \ref{sec:alignment}.}
    \label{fig:type2-gai-agent}
\end{figure}

Techniques like chain-of-thoughts \cite{wei2022chain} or tree-of-thoughts \cite{yao2024tree} can be adopted to prompt LLM to generate reasoning paths on-the-fly. GAI planning can effectively \emph{supplement} pre-defined reasoning for the long-tail behaviors and events. For further illustration, in Figure \ref{fig:type2-gai-agent} and the following paragraphs, we outline an interactive Recsys agent that enables seamless user engagements and active feedback.

Depending on the product and task requirements, there are practical considerations for setting up the action phase.
\begin{itemize}[leftmargin=*]
    \item \textbf{Database query and Recsys API call} --
    we discovered significant efficacy in developing LLM-oriented domain-specifc language (DSL) \cite{mernik2005and} that can be effectively translated into appropriate text2sql queries and API calls. When combined with standard few-shot learning and in-context learning/retrieval in run-time, the tool-using performance notably surpasses alternative options for getting data and recommendations.
    \item \textbf{LLM tools} -- to enable input augmentation, text summarization, and response generations, with "small" LLM such as \emph{Llama-7B} \cite{touvron2023llama} and \emph{Mistral-7B} \cite{jiang2023mistral} typically being sufficient for these tasks.
    \item \textbf{RAG tools} -- to facilitate the various Recsys curation functionalities described in Section \ref{sec:curation}.
    \item \textbf{Intent and ambiguity classification models} -- 
    ambiguities in user input can lead to failures in agents' planning process in practice. Using supplement models to characterize intents and ambiguity can enable agents to communicate with users for precise instructions.
    \item \textbf{Internal workflow and session data models} -- for real-time accessing and acting on the serving infrastructure such as data store (cache), schemas, streaming services, and more.
\end{itemize}

Lastly, the evaluation and continual optimization of Recsys agents involve more elements than what we outlined for RAG in Section \ref{sec:RAG-solution}. Many of the aspects require thorough examination, and we will elaborated them in the next sections.

\section{Responsible GAI and Human-AI Alignment}
\label{sec:alignment}
In social and e-commerce platforms, \emph{trust} and \emph{safety} stand out as the top product requirements for incorporating and serving LLM, RAG, and AI agents in Recsys. 
Subsequent changes are introduced to the offline evaluation and AI improvement strategies. In particular, \emph{risk assessment} of the generative outputs adds more dimensions to the traditional metrics like relevance, and \emph{aligning} GAI to user preferences also necessities a considerable expansion of the current serving and offline training strategies of Recsys.

\subsection{Trust and Responsible AI}
\label{sec:RAI}

Several identified attributes of GAI outputs can violate the trust and responsible AI principles, especially concerning \emph{bias} and \emph{toxicity} \cite{gehman2020realtoxicityprompts}, \emph{privacy} \cite{carlini2021extracting}, and \emph{hallucination} \cite{rawte2023survey}. It has also been found that LLM are subject to adversarial attacks with crafted inputs, e.g. \emph{jailbreak prompts}, to trigger undesired output \cite{zou2023universal}.

There are active investigations on the causes and modeling and data solutions of these issues \cite{schick2021self,dathathri2019plug, welbl2021challenges,wei2024jailbroken}. But as we mentioned in Section \ref{sec:introduction}, compound AI system has natural advantages in improving control and trust, and our tutorial will focus on introducing \textbf{system-level} solutions, especially:

\begin{itemize}[leftmargin=*]
    \item Adopting \textbf{Red Teaming}, which uses a "Red LLM" to regulate the behavior of the target LLM \cite{perez2022red}, has been demonstrated highly effective to alleviate both the bias and toxicity issues as well as preventing adversarial attacks and privacy leakage \cite{ganguli2022red,casper2306explore}.
    
    \item Incorporating system layers that use RAG (with knowledge base) or domain LLM and expert tools can often provide practical solutions to mitigate the hallucination problems \cite{zhang2023siren,tonmoy2024comprehensive}.  
\end{itemize}

\begin{figure}[htb]
    \centering
    \includegraphics[width=\linewidth]{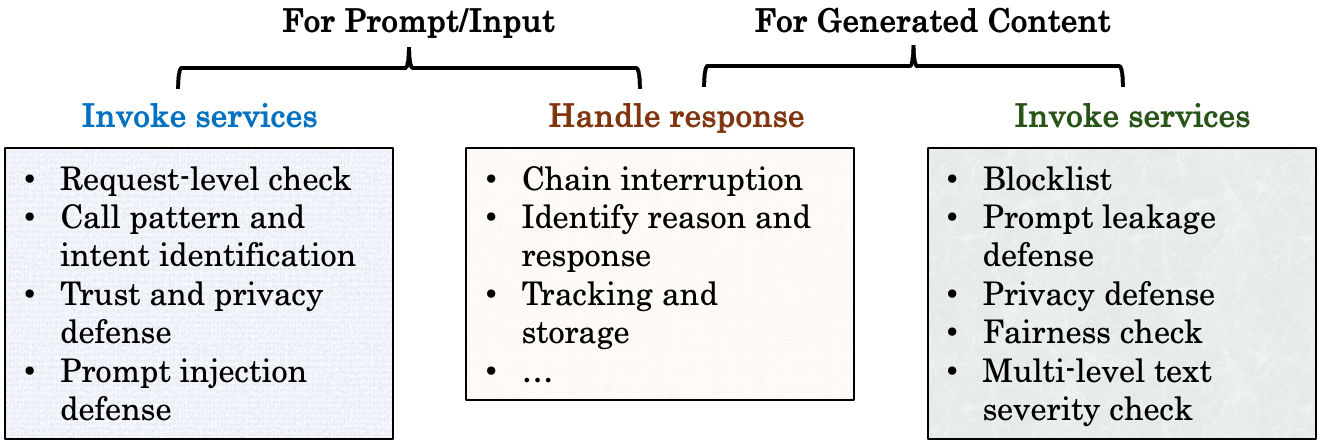}
    \caption{Core components of the Trust \& Responsible GAI service to support the various system-level solutions to bias, toxicity, privacy and adversarial attach issues.}
    \label{fig:RAI}
\end{figure}

In practice, the above solutions and related components constitute the \textbf{Trust \& Responsible GAI service}, which we elaborate in Figure \ref{fig:RAI}. After integrated with the existing API and environment manager, the service will apply to both the input and output stage of the generation process and is responsible for communicating with the clients through response handling.

\subsection{GAI response risk assessment}

Alongside the service-time guardrails solutions to ensure responsible GAI, offline risk assessment of the generated contents is also critical for safeguarding prompting, RAG, and agent solutions in Recsys by ensuring \emph{reliability} and \emph{trustworthiness} \cite{chang2023survey, zhang2023chatgpt}.

We specifically focus on assessing the risk perspective of GAI including ethical and bias issues and trust violations, given their crucial importance as product safeguards. Both conventional NLP testing approaches and innovative multifaceted exploration has been conducted to evaluate GAI’s toxicity, social bias, and trustworthiness vulnerabilities \cite{zhuo2023red, dhamala2021bold, gehman2020realtoxicityprompts, parrish2021bbq,wang2023decodingtrust}. 

\begin{figure}[hbt]
    \centering
    \includegraphics[width=\linewidth]{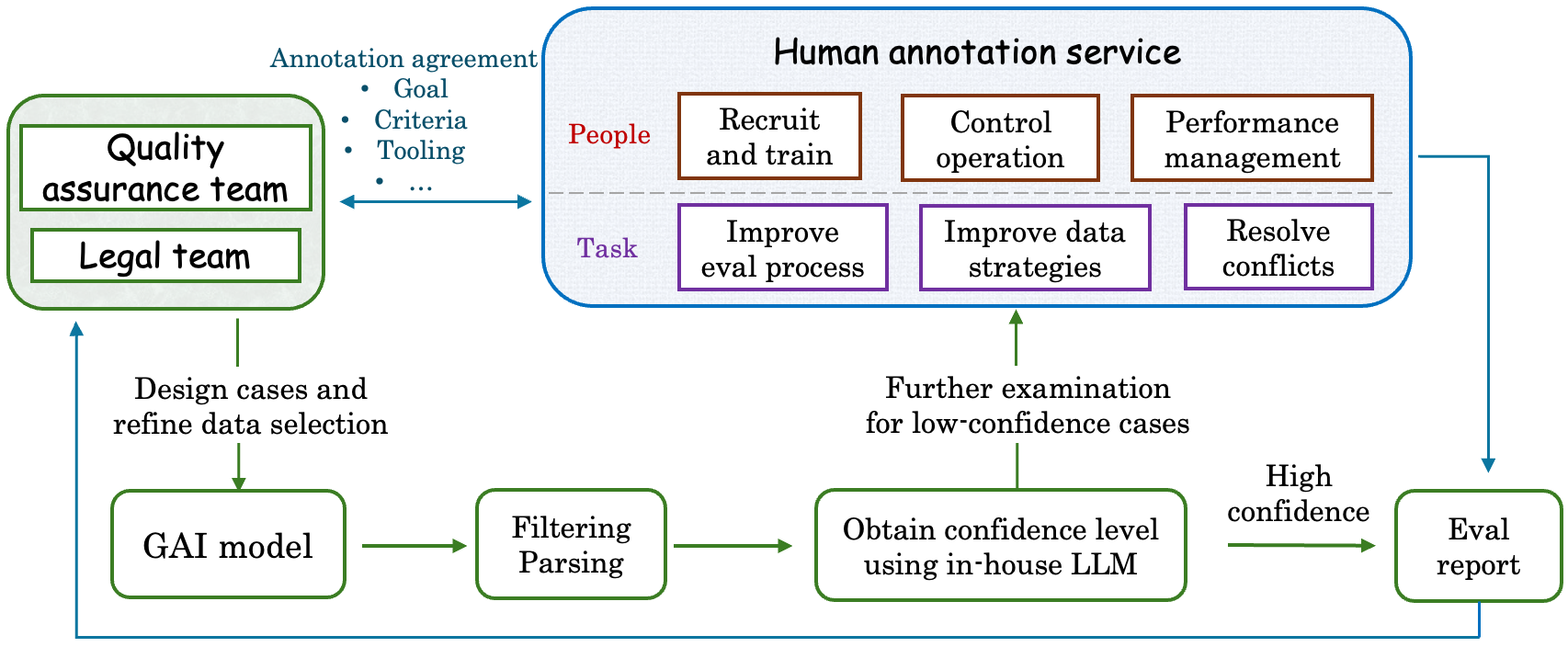}
    \caption{Workflow of human-in-the-loop response risk assessment for RAG and AI agent in Recsys. It can be applied to generic natural language generation evaluation \cite{howcroft2020twenty,zhou2022deconstructing}.}
    \label{fig:GET}
\end{figure}

While those research probe the problem from different angles, a common consensus underscores the necessity of \emph{dedicated} evaluation datasets and \emph{human-in-the-loop} assessments \cite{chang2023survey}. In Figure \ref{fig:GET}, we characterize a typical industrial service workflow for human-in-the-loop evaluation of GAI response risk assessment.

\subsection{Human-AI alignment for GAI in Recsys}

\begin{figure}[htb]
    \centering
    \includegraphics[width=0.8\linewidth]{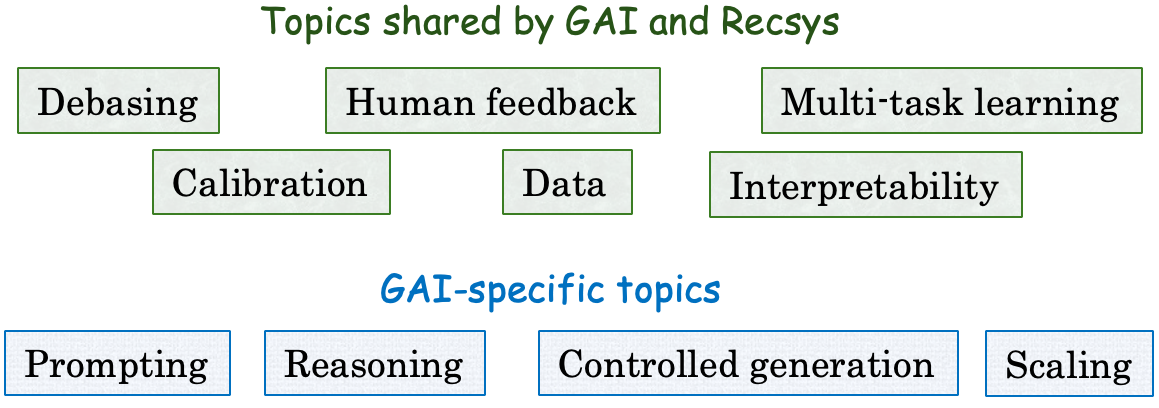}
    \caption{Some key directions for improving Human-AI alignment for GAI in Recsys.}
    \label{fig:alignment-topics}
\end{figure}

In the course to machine learning, AI alignment has shown up explicitly before in the field of Recsys and information retrieval where learning-to-rank from bandit feedback and user preference have been studied intensively \cite{li2010contextual,afsar2022reinforcement,liu2009learning,joachims2017unbiased}. As we summarized in Figure \ref{fig:alignment-topics}, there is non-trivial overlap between the Recsys and GAI methodological topics in the broader human-AI alignment domain. 

However, there are fundamental differences -- for GAI, alignment problems are generally refer to \emph{behavior alignment}, aiming create an agent that behaves in accordance with human instructions and values \cite{kenton2021alignment,ouyang2022training}. For Recsys, the alignment emphasizes more on matching and exploring human preferences and interests. 

Clearly, introducing new criteria, objectives, and techniques from GAI will be beneficial. For example, the \emph{Helpful, Honest, Harmless} criteria proposed by \citet{lin2021truthfulqa}, the adoption of \emph{reinforcement learning from human feedback} (RLHF) and \emph{direct preference optimization} (DPO) to achieve these criteria \cite{rafailov2024direct,bai2022training}, and the various technical directions outlined in \citet{bai2022training}. Another crucial aspect lies in \textbf{user interface design} and \textbf{communication strategies} \cite{weisz2024design}, as they can influence, regulate, incentivize, and direct users towards providing explicit preferences for continual learning through the aforementioned methods.

Understanding the cause of \emph{misalignment} in GAI can also go a long way. Specifically, the emerging GAI capabilities are often attributed to the scaling law and extensive pre-training corpus \cite{kaplan2020scaling,bubeck2023sparks}, so building insights to the data and model scaling properties can be critical for diagnosing issues and identifying opportunities \cite{bender2021dangers,dodge2021documenting,henighan2020scaling}. A comprehensive solution can therefore incorporate both the traditional wisdom of \emph{data cleaning}, \emph{debiasing}, \emph{multi-task learning}, as well as contemporary GAI techniques like calibration of prompting \cite{zhao2021calibrate} and reasoning with self-consistency \cite{wang2022self}. Additionally, enforcing \emph{controlled generation} by such as Red Teaming is also a practical approach for human-AI alignment.

\section{Open Problems}
\label{sec:open-problem}
In the course of landing GAI in Recsys, we have faced several universal challenges on top of the application-specific problems outlined in the previous sections. Here, we emphasize a selection of these that hold particular practical significance and opportunity.

\textbf{Effective in-house GAI serving stack} -- for industrial systems, GAI \emph{cost} and \emph{latency} (especially those from the long input and token generation) significantly impact the experience of both developers and users when adopting GAI at the scale of Recsys, and within the constraints of existing service-level agreements. \cite{openai2023slow,pope2023efficiently,chen2023frugalgpt,mlops2023responses}. Developing a systematic solution remains challenging despite the recent research advances in LLM serving \cite{miao2023towards}.

\textbf{Imitating LLM on proprietary data and models} -- constructing compact models that strikes a balance between \emph{task imitation} with proprietary model and structured data, alongside LLM's \emph{generation capability}, can be instrumental for overcoming the computational and resource challenges while ensuring optimal efficiency and effectiveness for specific applications \cite{gudibande2023false,chung2022scaling,kaddour2023challenges}. 

\textbf{High-quality human data}
-- high quality data is the fuel for all stages in the GAI and Recsys development cycles. Unlike Recsys which can leverage a wide range of user feedback, the majority of task-specific labeled data for GAI training and evaluation is obtained through \emph{human annotation}, including tasks such as classification or RLHF labeling for LLM alignment training. While numerous machine learning techniques can contribute to enhancing data quality, human data collection fundamentally relies on attention to detail and careful execution \cite{callison2009fast, davani2022deal, aroyo2015truth, pleiss2020identify}.

\textbf{Online evaluation and monitoring for GAI in Recsys} -- while the Recsys community has developed mature and sophisticated solutions for online controlled experiments \cite{kohavi2020trustworthy,kohavi2007practical,kreuzberger2023machine}, with the introduction of GAI, measuring online performance is confronted with challenges such as overcoming brittle metrics, measuring generative task performance at scale, ensuring reproducibity, and coming up with the appropriate monitoring and experimental designs for both user satisfaction and system efficiency \cite{chen2023chatgpt,liang2022holistic,lu2021fantastically,srivastava2022beyond}. 


\section{Conclusion}
\label{sec:conclusion}

In this survey, we provide a comprehensive review of the promising initial attempts at integrating GAI into social and e-commerce Recsys. We traverse the landscape of industrial Recsys and GAI fundamentals, existing solution frameworks, their connections to research advancements, and the practical insights and challenges encountered. As a pioneering contribution in this field, our survey also serves as a guide for incorporating GAI into a broader range of industrial Recsys. Moreover, it benefits scholars and practitioners in both the GAI and Recsys communities by delineating clear practical roadmaps for embedding GAI into real-world applications.

\bibliographystyle{ACM-Reference-Format}
\bibliography{paper-references}

\end{document}